\documentclass{article}
\usepackage{amsmath}[1996/11/01]
\usepackage{amssymb,amsthm,amsxtra}
\usepackage{epsfig}

\newlength{\mytopmargin}
\newlength{\myleftmargin}
\def\zz{\rlx\hbox{\small \sf Z\kern-.4em Z}}

\newtheorem{lemma}{Lemma}[section]

\newtheorem{prop}[lemma]{Proposition}

\begin{document}

\vspace{1cm}
\noindent
\begin{center}{   \large \bf Quantum conductance problems and the Jacobi ensemble } 
\end{center}
\vspace{5mm}

\noindent
\begin{center}
 P.J.~Forrester\\

\it Department of Mathematics and Statistics, \\
University of Melbourne, Victoria
3010, Australia
\end{center}
\vspace{.5cm}
\begin{quote}
In one dimensional transport problems the scattering matrix $S$ is decomposed into a block
structure corresponding to reflection and transmission matrices at the two ends. For
$S$ a random unitary matrix, the singular value probability distribution function of
these blocks is calculated. The same is done when $S$ is constrained to be symmetric,
or to be self dual quaternion real, or when $S$ has real elements, or has real
quaternion elements. Three methods are used: metric forms; a variant of the Ingham-Seigel
matrix integral; and a theorem specifying the Jacobi random matrix ensemble in terms of
Wishart distributed matrices.
\end{quote}

\vspace{.5cm}
\noindent
\section{Introduction}
\setcounter{equation}{0}
In mesoscopic physics, the conductance of certain quasi one-dimensional wires containing
scattering impurities exhibit the phenomenum known as universal conductance fluctuations.
As first predicted theoretically \cite{Al85,LS85}, and soon after observed experimentally
\cite{WW86}, for such wires the variance of the conductance is of order $(e^2/h)^2$,
independent of sample size or disorder strength. Furthermore the variance decreases by precisely
a factor of two if time reversal symmetry is broken by a magnetic field. This conductance
problem is fundamental for its relation to the Landauer scattering theory of electronic
conduction (see \cite{Be97} and references therein), and to time reversal symmetry.
It is further fundamental for its relation to random matrix theory. Let us revise these points by way
of background and motivation for the specific problem of this paper.

In the theoretical description of the conductance problem, basic quantities are the electron fluxes
at the left and right hand edges of the wire. These are specified by an $n$-component vector
$\vec{I}$ and an $m$-component vector $\vec{I}'$ specifying the complex amplitudes of the
available plane wave states travelling into the left and right sides of the wire respectively,
as well as an $n$-component vector $\vec{O}$ and an $m$-component vector $\vec{O}'$ for the
same states travelling out of the left and right sides of the wire. For definiteness it will
be assumed that $n \ge m$. 

By definition, the scattering matrix $S$ relates the flux
travelling into the conductor to that travelling out,
\begin{equation}
 S \left [ \vec {I} \atop \vec {I'} \right ] :=
\left [ \vec {O} \atop \vec {O'} \right ].
\label{2.1}
\end{equation}
The scattering matrix is further decomposed in terms of reflection
and transmission matrices by
\begin{equation}
 S = \left [ \begin{array}{cc}
 r_{n \times n} &  t'_{n \times m} \\
 t_{m \times n} &  r'_{m \times m} \end{array} \right ].
\label{2.4aa}
\end{equation}
Flux conservation requires
$$
|\vec {I}|^2 + |\vec{I'}|^2 = |\vec {O}|^2 + |\vec {O'}|^2,
$$
and this implies that $S$ must be unitary. Furthermore, by relating $S$ to an evolution operator
and thus a Hamiltonian, one can argue (see e.g.~\cite{Fo02}) that $S$ must be symmetric if the
system has a time reversal symmetry with $T^2=1$, and a self dual quanternion matrix when there is
a time reversal symmetry with $T^2=-1$.

The immediate relevance of the above formalism is seen by invoking the 
 Landauer scattering theory of electronic
conduction. According to this formalism, the conductance $G$ is given in terms of the
transmission matrix $t_{m \times n}$ (or $ t'_{n \times m}$) by the so called two probe
 Landauer formula
\begin{equation}
G / G_0 = {\rm Tr} ( t^\dagger  t) = {\rm Tr} ( t'^\dagger  t')
\label{2.6ab}
\end{equation}
where $G_0 = 2 e^2 / h$ is twice the fundamental quantum unit of conductance.
Thus to compute $G$ it suffices to know the distribution of the eigenvalues of
$t^\dagger t$ (or $t'^\dagger t'$). In fact the matrix
$S$ can be decomposed in a form which isolates these eigenvalues.

For definiteness suppose there is no time reversal symmetry, so $S$ is a general
unitary matrix. Each block of $S$ can then be decomposed according to a general
singular value decomposition. For example, for the block $r_{n \times n}$ we have
$$
r_{n \times n} = U_r \Lambda_r V_r^\dagger,
$$
where $U_r$, $V_r$ are unitary matrices and $\Lambda_r$ is a rectangular diagonal matrix with
entries equal to the positive square roots of the eigenvalues of $r^\dagger r$ (these
eigenvalues are between 0 and 1 since $r^\dagger r + t^\dagger t = I_{n \times n}$). The
unitarity of $S$ inter-relates the matrices $U_r, U_{r'}, \dots$ (see e.g.~\cite{SMMP91}) and
implies the decomposition
\begin{equation}\label{2.91x}
 S =
\left [ \begin{array}{cc}
 U_{r} & { 0} \\ { 0} &  U_{r'}  \end{array}\right ]  L
\left [ \begin{array}{cc}
 V_{r}^\dagger & { 0} \\ { 0} &  V_{r'}^\dagger
\end{array} \right ]
\end{equation}
where
\begin{equation}
 L := \left [ \begin{array}{cc}
 \sqrt{1 -  \Lambda_t  \Lambda_t^T} & i  \Lambda_t \\
 i \Lambda_t^T &\sqrt{{ 1} -  \Lambda_t^T  \Lambda_t}
\end{array}\right ].
\label{2.5f}
\end{equation} 

Symmetries relate $V_r,V_{r'}$ to $U_r, U_{r'}$. As already remarked, in the case of a time
reversal symmetry $T^2 = 1$, $S$ must be symmetric, while for $T^2 = -1$, $S$ must be self dual
quaternion real. These symmetries require that
\begin{equation}\label{2}
V_r^\dagger = U_r^T, \, V_{r'}^\dagger = U_{r'}^T \qquad
{\rm and} \qquad 
V_r^\dagger = U_r^D, \, V_{r'}^\dagger = U_{r'}^D
\end{equation}
respectively. In (\ref{2}) the operation $D$, for an $n \times n$ matrix $A$ with real
quaternion elements regarded as a $2n \times 2n$ matrix with complex elements, is specified
by
$$
A^D = Z_{2n} A^T Z_{2n}^{-1}, \qquad Z_{2n} := {I_n} \otimes 
\left [ \begin{array}{cc} 0 & -1 \\ 1 & 0 \end{array} \right ].
$$

The relevance of random matrix theory comes from hypothesizing that apart from the constraint
imposed by a time reversal symmetry, in the regime relevant to universal conductance
fluctuations, the scattering matrix $S$ is effectively a random matrix chosen with Haar
(uniform) measure. With this assumption the non-zero elements of $\Lambda_t$, which in
turn are equal to the square root of the non-zero eigenvalues of $t^\dagger t$ and are thus
in the interval $(0,1)$, has a joint probability density function  (p.d.f.) proportional to
\begin{equation}\label{2.1b}
\prod_{j=1}^m\lambda_j^{\beta \alpha}
\prod_{1 \le j < k \le m} |\lambda_k^2 - \lambda_j^2|^\beta, \qquad
\alpha := n - m + 1 - 1/\beta,
\end{equation}
where $\beta = 1, 2, 4$ according to $S$ being constrained to be symmetric, no constraints,
or constrained to be self dual quanternion.
In its full generality this result is stated without proof in Beenakker \cite[eq.~(2.9)]{Be97},
where it is attributed to Brouwer. In the case $n=m$ it was derived in
\cite{BM94, JPB94}.
With knowledge of (\ref{2.1b}), it can be deduced from (\ref{2.6ab}) that with $\alpha$ fixed
\cite{Be93a}
\begin{equation}
\lim_{m,n \to \infty} {\rm Var}(G/G_0) = {1 \over 8 \beta},
\end{equation}
which is in quantitative agreement with experiment. 

However the issue of fluctuation formulas is not our concern here.
Rather we seek to specify three distinct derivations of (\ref{2.1b}) in the case
$\beta = 2$,  each of which we believe
to have particular features of interest.
The first derivation to be given uses the method of metric forms. 
This method applies to all three cases, and in fact details will be given in the
case $\beta = 1$. 
The second derivation uses versions of
the Ingham-Seigel integral; while the third relies on knowledge of the 
Jacobi random matrix ensemble as derived from Wishart distributed matrices.
These latter two derivations generalize (\ref{2.1b}) in the case
$\beta = 2$, and also allow $\beta$ generalizations
relating to the block decomposition of unitary matrices, however now with
$\beta = 1$ refering to the elements of $S$ being real, and $\beta = 4$ to the
elements being real quaternion. Some aspects of statistical properties of the
block decomposition of real orthogonal matrices (i.e.~unitary matrices with real
elements) have been previously given in \cite{ZS99}.

\section{Metric forms}
\setcounter{equation}{0}
Let $X = [x_{jk}]_{j,k=1,\dots,N}$ be an $N \times N$ matrix. Let $\{x_\mu\}$, where $\mu$ labels
both rows and columns, be the set of independent real and imaginary parts in $X$. The metric
form of the line element $ds$ is defined by
$$
(ds)^2 := {\rm Tr}( d X d X^\dagger) =
\sum_{\mu \: {\rm diag.}} (d x_\mu)^2 + 2
\sum_{\mu \: {\rm upper} \atop {\rm triangular} }
(d x_\mu)^2.
$$
The corresponding volume measure is
\begin{equation}
(d X) = \bigwedge_{\mu \: {\rm diag}.} d x_\mu
\bigwedge_{\mu \: {\rm upper} \atop {\rm triangular} } dx_\mu.
\end{equation}
Suppose now a change of variables $\{x_{\mu} \} \mapsto \{ y_\mu \}$ is made such that $(ds)^2$ is
a symmetric quadratic form in $\{ d y_\mu \}$,
$$
( ds)^2 = \sum_{\mu, \nu} g_{\mu, \nu} dy_\mu dy_\nu, \qquad
g_{\mu,\nu} = g_{\nu,\mu}.
$$
The corresponding volume measure is then (see e.g.~\cite{MF53})
\begin{equation}\label{1.shdy}
(\det [ g_{\mu, \nu} ] )^{1/2} \bigwedge_\mu dy_\mu.
\end{equation}

In \cite{Fo02} this formalism has been used to derive the result (\ref{2.1b}) in the case
$\beta = 2$. Here the details will be given for the case $\beta = 1$, when (\ref{2.91x}) reads
\begin{equation}\label{s1}
 S =
\left [ \begin{array}{cc}
 U_{r} & { 0} \\ { 0} &  U_{r'}  \end{array}\right ]  L
\left [ \begin{array}{cc}
 U_{r}^T & { 0} \\ { 0} &  U_{r'}^T
\end{array} \right ],
\end{equation}
will be given.

For a general matrix $A$, let $\delta A := A^\dagger d A$, where $d A$ denotes the matrix of
differentials of the elements of $A$. Using the fact that for $A$ unitary
$(\delta A )^\dagger = - \delta A$, it follows from (\ref{s1}) that
\begin{equation}\label{s2}
\left [ \begin{array}{cc}
 U_{r}^\dagger & { 0} \\ { 0} &  U_{r'}^\dagger  \end{array}\right ]
d S
\left [ \begin{array}{cc}
 \bar{U}_{r} & { 0} \\ { 0} &  \bar{U}_{r'}  \end{array}\right ]
=
\left [ \begin{array}{cc}
 \delta U_{r} & { 0} \\ { 0} &  \delta U_{r'}  \end{array}\right ]  L + d L -
L \left [ \begin{array}{cc}
 \delta \bar{U}_{r} & { 0} \\ { 0} &  \delta \bar{U}_{r'}  \end{array}\right ].
\end{equation}
In Tr$(d S d S^\dagger)$, the right hand side of (\ref{s2}) can effectively be substituted for
$d S$. Doing this and simplifying using
$$
{\rm Tr} \left (
\left [ \begin{array}{cc}
 \delta U_{r} & { 0} \\ { 0} &  \delta U_{r'}  \end{array}\right ] L^\dagger d L
\right ) = 0
$$
gives
\begin{equation}\label{s3}
{\rm Tr}(d S d S^\dagger) = {\rm Tr} \, A_1 \bar{A}_1 +  {\rm Tr} (A_2 + \bar{A}_2) +
{\rm Tr}(A_3 \bar{A}_3) + {\rm Tr}( dL d L^\dagger) 
\end{equation}
where, with $I$ denoting the identity matrix,
\begin{eqnarray*}
A_1 & = & \sqrt{I - \Lambda_t \Lambda_t^T} \delta \bar{U}_r -
\delta U_r \sqrt{I - \Lambda_t \Lambda_t^T}  \\
A_2 & = & ( \Lambda_t  \delta \bar{U}_{r'} - \delta U_r \Lambda_t )
( \Lambda_t^T  \delta {U}_{r} - \delta \bar{U}_{r'} \Lambda_t^T ) \\
A_3 & = &  \sqrt{I - \Lambda_t \Lambda_t^T} \delta \bar{U}_{r'} -
\delta U_{r'} \sqrt{I - \Lambda_t \Lambda_t^T}.
\end{eqnarray*}

We recall that the diagonal elements of $\Lambda_t$ are the positive square roots of the
eigenvalues of $t^\dagger t$. Because $t = t_{m \times n}$ and thus has rank $m$, 
$t^\dagger t$ must have
$n - m$ zero eigenvalues, so with the diagonal elements of $\Lambda_t\Lambda_t^T$ denoted $\lambda_j$
$(j=1,\dots,n)$, 
as is consistent with the notation of (\ref{2.1b}),
we have $\lambda_{m+1} = \cdots = \lambda_n = 0$. Using this fact we see that
in component form the four terms in (\ref{s3}), $T_1,\dots,T_4$ say, can be expanded to read
\begin{eqnarray*}
T_1 & = & \sum_{k=1}^n(1 - \lambda_k^2)
|(\delta \bar{U}_r)_{kk} - (\delta U_r)_{kk}|^2
+ \sum_{1 \le k < l \le n} {1 \over 2}
\Big (\sqrt{1 - \lambda_l^2} + \sqrt{1 - \lambda_k^2} \Big )^2
|(\delta \bar{U}_r)_{lk} - (\delta U_r)_{lk}|^2 \nonumber \\
&& + \Big (
\sum_{1 \le k < l \le m} + \sum_{k=1}^m \sum_{l=m+1}^n \Big )
{1 \over 2}
\Big (\sqrt{1 - \lambda_l^2} - \sqrt{1 - \lambda_k^2} \Big )^2
|(\delta \bar{U}_r)_{lk} + (\delta U_r)_{lk}|^2
\label{T1}
\\
T_2  & = & 2 \bigg (\sum_{k=1}^m \lambda_k^2
|(\delta \bar{U}_{r'})_{kk} - (\delta  U_r)_{kk}|^2 +
\sum_{1 \le k < l \le m} \Big \{ {1 \over 2}(\lambda_l + \lambda_k)^2
|(\delta \bar{U}_{r'})_{lk} - (\delta  U_r)_{lk}|^2  \nonumber \\
&& + {1 \over 2}(\lambda_l - \lambda_k)^2
|(\delta \bar{U}_{r'})_{lk} + (\delta U_r)_{lk}|^2 \Big \} +
\sum_{k=1}^m \sum_{l=m+1}^n \lambda_k^2 |(\delta  U_r)_{lk}|^2 \bigg )
\label{T2}
\\
T_3 & = & \sum_{k=1}^m(1 - \lambda_k^2)
|(\delta \bar{U}_{r'})_{kk} - (\delta U_{r'})_{kk}|^2
+ \sum_{1 \le k < l \le m} {1 \over 2}
\Big (\sqrt{1 - \lambda_l^2} + \sqrt{1 - \lambda_k^2} \Big )^2
|(\delta \bar{U}_{r'})_{lk} - (\delta  U_{r'})_{lk}|^2 \nonumber \\
&& +
{1 \over 2}
\Big (\sqrt{1 - \lambda_l^2} - \sqrt{1 - \lambda_k^2} \Big )^2
|(\delta \bar{U}_{r'})_{lk} + (\delta U_{r'})_{lk}|^2
\label{T4}
\\
T_4 & = & 2 \sum_{j=1}^m {(d \lambda_j)^2 \over 1 - \lambda_j^2}.
\label{T5}
\end{eqnarray*}

In general a symmetric unitary matrix has the same number of independent elements as a real
symmetric matrix of the same rank, so $S$ has ${1 \over 2}(m+n)(m+n+1)$ independent elements.
We thus seek this same number of independent differentials in $T_1,\dots,T_4$. These are
$$
(\delta U_r)_{kk}^{(\rm i)}, \quad 1 \le k \le n \qquad
(\delta U_{r'})_{kk}^{(\rm i)}, \quad 1 \le k \le m
$$
and
$$
(\delta U_r)_{lk}^{(\rm r)}, \quad (\delta U_r)_{lk}^{(\rm i)}, \quad
(\delta U_r)_{lk}^{(\rm r)}, \quad (\delta U_r)_{lk}^{(\rm i)}, \quad
1 \le k < l \le m, 
$$
as well as
$$
(\delta U_r)_{lk}^{(\rm r)}, \quad (\delta U_r)_{lk}^{(\rm i)}, \qquad
1 \le k \le m \: \& \: m+1 \le l \le n
$$
and
$$
 (\delta U_r)_{lk}^{(\rm i)}, \quad  m+1 \le l < k \le n  \qquad
d \lambda_j \quad 1 \le j \le m,
$$
which indeed tally to ${1 \over 2}(m+n)(m+n+1)$.

We see from the expressions for $T_1, T_2, T_3$ that the contribution to the metric form 
from the differentials with subscripts $lk$ such that $1 \le k < l \le m$ is
\begin{eqnarray}
&&4 \sum_{1 \le k < l \le m} \Big ( 1 + \sqrt{1 - \lambda_l^2} \sqrt{1 - \lambda_k^2} \Big )
\Big ( (\delta U_r^{(\rm i)})_{lk}^2 + (\delta U_{r'}^{(\rm i)})_{lk}^2 \Big )
\nonumber \\
&& \qquad +
\Big ( 1 - \sqrt{1 - \lambda_l^2} \sqrt{1 - \lambda_k^2} \Big )
\Big ( (\delta U_r^{(\rm r)})_{lk}^2 + (\delta U_{r'}^{(\rm r)})_{lk}^2 \Big )
\nonumber \\
&&
\qquad  - 2 \lambda_l \lambda_k  (\delta U_{r'}^{(\rm r)})_{lk} (\delta U_r^{(\rm r)})_{lk}
+ 2  \lambda_l \lambda_k  (\delta U_{r'}^{(\rm i)})_{lk} (\delta U_r^{(\rm i)})_{lk}. 
\end{eqnarray}
Setting $\sqrt{1 - \lambda_l^2} \sqrt{1 - \lambda_k^2} =:a$ for notational convenience, this
portion of the metric form
contributes to  $(\det[g_{jk}])^{1/2}$ in (\ref{1.shdy}) $2 \times 2$ block factors which is
proportional to
\begin{equation}\label{f1}
\prod_{k<l}^m \left | \begin{array}{cc} 1+ a & - \lambda_l \lambda_k \\
- \lambda_l \lambda_k & 1 + a \end{array} \right |^{1/2}
\left | \begin{array}{cc} 1- a &  \lambda_l \lambda_k \\
 \lambda_l \lambda_k & 1 - a \end{array} \right |^{1/2} = \prod_{k<l}^m |\lambda_l^2 - \lambda_k^2|.
\end{equation}

For $m+1 \le k < l \le n$, the coefficient of $(\delta U_r)_{lk}^{({\rm i})}$ in 
$T_1$ is independent of the
$\lambda$'s and so for the present purposes can be ignored. For $1 \le k \le m$, $m+1 \le l \le n$, we
see from $T_1$ and $T_2$ that the corresponding contribution to the metric form is
$$
4 \sum_{k=1}^m \sum_{l=m+1}^n (1 + \sqrt{1 - \lambda_k^2}) 
((\delta U_r)_{lk}^{({\rm i})})^2 +
(1 - \sqrt{1 - \lambda_k^2}) ((\delta U_r)_{lk}^{({\rm r})})^2.
$$
This contributes to the volume form a factor proportional to
\begin{equation}\label{f2}
\prod_{k=1}^m \prod_{l=m+1}^n \lambda_k = \prod_{k=1}^m \lambda_k^{n-m}.
\end{equation}

We read off that the contribution to $(\det[g_{jk}])^{1/2}$ from the coefficients of the terms 
$(d \lambda_j)^2$ in $T_4$ is proportional to
\begin{equation}\label{f3}
\prod_{k=1}^m {1 \over (1 - \lambda_k^2)^{1/2} }.
\end{equation}

It remains to calculate the contribution from the differentials on the diagonal. For $m+1 \le k \le
n$, the coefficient of $(\delta U_r)_{kk}^{(i)}$ is a constant so these differentials can be
ignored. We read off from $T_1, T_2, T_3$ that the contribution to the metric form from the
remaining differentials is
$$
2 \sum_{k=1}^m (2 - \lambda_k^2) \Big (
((\delta U_r)_{kk}^{({\rm i})})^2 + (\delta U_{r'})_{kk}^{({\rm i})})^2  \Big ) +
2 \lambda_k^2 (\delta U_r)_{kk}^{({\rm i})} (\delta U_{r'})_{kk}^{({\rm i})}.
$$
The contribution to the volume form is thus proportional to
\begin{equation}\label{f4}
\prod_{k=1}^m \left | \begin{array}{cc} 2 - \lambda_k^2 & \lambda_k^2 \\
\lambda_k^2 & 2 - \lambda_k^2 \end{array} \right |^{1/2} \: \propto \:
\prod_{k=1}^m (1 - \lambda_k^2)^{1/2}.
\end{equation}

Multiplying together (\ref{f1})--(\ref{f4}) gives (\ref{2.1b}) in the case $\beta = 1$.

\section{Matrix integrals}
\setcounter{equation}{0}
Let
\begin{equation}\label{F0}
I_{m,n}^{(2)}(Q_m) := \int e^{{i \over 2} {\rm Tr}(H_m Q_m)}
\Big ( \det (H_m - \mu I_m) \Big )^{-n} (d H_m)
\end{equation}
where $H_m, Q_m$ are $m \times m$ Hermitian matrices, and suppose $n \ge m$,
${\rm Im} \, \mu > 0$. For $Q_m$ positive definite, it has been proved by
Fyodorov \cite{Fy02} that
\begin{equation}\label{F1}
I_{m,n}^{(2)}(Q_m) =
{2^m \pi^{m(m+1)/2} i^m (-1)^{m(m-1)/2} \over
\prod_{j=0}^{m-1} \Gamma(n-j) }
\Big ( \det ( {i \over 2} Q_m) \Big )^{n-m} e^{{i \over 2} \mu {\rm Tr} \, Q_m}.
\end{equation}
(Here we have taken $(dH_m) := \prod_{j=1}^m dh_{jj}^{(\rm i)}
\prod_{j < k} dh_{jk}^{(\rm r)}  dh_{jk}^{(\rm i)}$ which differs by a factor of 2
in the product over $j < k$ to the convention adopted in \cite{Fy02}.) 
This matrix integral may be regarded as being of the type first evaluated
by Ingham and Siegel (see \cite{Fy02} and references therein). In a subsequent
work \cite{FS03}, (\ref{F0}) was used to derive (\ref{2.1b}) in the case
$\beta = 2$. Here we will show that 
this derivation can be used to derive a generalization of (\ref{2.1b}) in the
case $\beta = 2$, and this generalization can be further extended to 
the cases $\beta = 1$ and
$\beta = 4$, using  suitable variants of (\ref{F0}), where now $\beta = 1$ and
4 refers to
the decomposition (\ref{2.4aa}) with $S$ having real and real quaternion elements
respectively.

Define by $I_{m,n}^{(1)}(Q_m)$ the matrix integral (\ref{F0}) with $H_m$ and
$Q_m$ now real symmetric. Also, define by $I_{m,n}^{(4)}(Q_m)$ the same matrix
integral but with $H_m$ and $Q_m$ now self dual quaterion Hermitian matrices (such
matrices regarded as $2m \times 2m$ Hermitian matrices are doubly degenerate;
adopt the convention that the operations Tr and det include only distinct
eigenvalues). In \cite{Fy02} it is remarked that the method of derivation given
therein to deduce that the evaluation of $I_{m,n}^{(2)}(Q_m)$ can also be used
to deduce the evaluation of $I_{m,n}^{(1)}(Q_m)$, which reads
\begin{equation}\label{F2}
I_{m,n/2}^{(1)}(Q_m) =
{2^m \pi^{m(m+3)/2} i^{m(m+1)/2}  \over
\prod_{j=0}^{m-1} \Gamma((n-j)/2) }
\Big ( \det ( {i \over 2} Q_m) \Big )^{(n-m-1)/2} e^{{i \over 2} \mu {\rm Tr} \, Q_m}.
\end{equation}
Applying the same method to $I_{m,n}^{(4)}(Q_m)$ gives 
\begin{equation}\label{F3}
I_{m,2n}^{(4)}(Q_m) =
{(2i)^m \pi^{m^2} \over \prod_{j=0}^{m-1} \Gamma(2(n-j)) }
\Big ( \det ( {i \over 2} Q_m) \Big )^{2(n-m+1/2)}
e^{{i \over 2} \mu {\rm Tr} \, Q_m}.
\end{equation}
Hence, all three cases we have
\begin{equation}\label{F4}
I_{m,\beta n/2}^{(\beta)}(Q_m) =
C_{m,n}^{(\beta)}
( \det (  Q_m) )^{(\beta / 2) (n-m+1-2/\beta)}
e^{{i \over 2} \mu {\rm Tr} \, Q_m},
\end{equation}
where $C_{m,n}^{(\beta)}$ is independent of $Q_m$.

Let $U$ be an $N \times N$ random unitary matrix, with real ($\beta = 1$),
complex $(\beta = 2$) and real quaternion ($\beta = 4$), chosen with Haar
measure. Generalizing (\ref{2.4aa}), decompose $U$ into blocks
\begin{equation}\label{2.4bb}
U = \left [ \begin{array}{cc} A_{n_1 \times n_2} & C_{n_1 \times (N - n_2)} \\
B_{(N-n_1) \times n_2} & D_{(N-n_1) \times (N - n_2)} \end{array} \right ].
\end{equation}
Since $U$ is unitary, we require
\begin{equation}\label{AC}
A A^\dagger + C C^\dagger = I_{n_1},
\end{equation}
together with three similar equations involving $B$ and $D$
which given $A$ and $C$ can always be
satisfied. The idea of \cite{ZS99} is to regard (\ref{AC}) as a constraint in the
space of general rectangular matrices $A,C$ with entries of the type required by the
index $\beta$. Thus in this viewpoint the distribution of $A$ is given by
\begin{equation}\label{AC1}
\int \delta (A A^\dagger + C C^\dagger - I_{n_2}) (d C).
\end{equation}

In (\ref{AC1}) the delta function is a product of scalar delta functions, one for each
independent real and imaginary component of $A^\dagger A + C^\dagger C - I_{n_2}$.
It is proportional to the matrix integral
\begin{equation}\label{AC2}
\int e^{- i {\rm Tr} ( H (A A^\dagger + C C^\dagger - I_{n_2}))} (d H)
\end{equation}
where $H$ is an Hermitian matrix with elements of the type $\beta$. Following
\cite{FS03} we would like to substitute (\ref{AC2}) for the delta function in
(\ref{AC1}), and change the order of integration. The integrations over $(dC)$
are simply Gaussian integrals. For the resulting function of $H$ to be integrable
around $H=0$, the replacement $H \mapsto H - i \mu I_{n_1}$ in the exponent of
(\ref{AC2}) must made. Doing this and computing the Gaussian integrals gives that
(\ref{AC1}) is proportional to
$$
\lim_{\mu \to 0^+} \int \Big ( \det ( H - i \mu I_{n_1} ) \Big )^{-\beta (N -
n_2)/2} e^{i {\rm Tr}(H(I_{n_1} - A A^\dagger))} (dH)
$$
Evaluating the matrix integral using (\ref{F4}) shows that the distribution of $A$
is proportional to
\begin{equation}\label{AC3}
\Big ( \det ( I_{n_1} - A A^\dagger ) \Big )^{(\beta/2)(N - n_1 - n_2 + 1 - 2/\beta))}.
\end{equation}
Note that for this to be normalizable, we must have
\begin{equation}\label{Nn}
N - n_1 - n_2 \ge 0.
\end{equation} 

Suppose that in addition to (\ref{Nn}) we have
$n_1 \ge n_2$. Then $A$ has rank $n_1$ and so $A A^\dagger $ has
$n_1 - n_2$ zero eigenvalues and (\ref{AC3}) can be written
\begin{equation}\label{AC3a}
\Big ( \det ( I_{n_2} - A^\dagger A) \Big )^{(\beta/2)(N - n_1 - n_2 + 1 - 2/\beta)}.
\end{equation}
Setting
$Y = A^\dagger A$ we know (see e.g.~\cite{Fo02}) that
\begin{equation}\label{AC3b}
d A \propto (\det Y)^{(\beta/2)(n_1 - n_2 + 1 - 2/\beta)} (d Y),
\end{equation}
so we have from (\ref{AC3a}) that the distribution of $Y$ is proportional to
\begin{equation}\label{AC4}
(\det Y)^{(\beta/2)(n_1 - n_2 + 1 - 2/\beta))}
\Big ( \det ( I_{n_2} - Y) \Big )^{(\beta/2)(N - n_1 - n_2 + 1 - 2/\beta)}.
\end{equation}
Denote the eigenvalues of $Y$ by $y_1,\dots, y_{n_2}$. Using the fact that the
eigenvalue dependent portion of the Jacobian for an Hermitian matrix, with elements
of the type $\beta$, when changing variables to its eigenvalues and eigenvectors is
$\prod_{j < k} |y_k - y_j|^\beta$, we read off from (\ref{AC4}) that the eigenvalue
p.d.f.~of $Y$ is proportional to
\begin{equation}\label{AC5}
\prod_{j=1}^{n_2} y_j^{(\beta/2)(n_1 - n_2 + 1 - 2/\beta)}
(1 - y_j)^{(\beta/2)(N - n_1 - n_2 + 1 - 2/\beta))}
\prod_{j < k}^{n_2} |y_k - y_j|^\beta .
\end{equation}
In the case 
\begin{equation}\label{3.25a}
n_1 = n_2 = m, \qquad N = n+m, 
\end{equation}
the matrix $A$ in (\ref{2.4bb})
coincides with $r$ in (\ref{2.4aa}), and the $y_j$ in
(\ref{AC5}) are then the eigenvalues of $r^\dagger r$. 
The non-zero values of the singular values
of the submatrix  $t$ in (\ref{2.4aa}) are the $\lambda_j$'s in (\ref{2.1b}).
The decomposition (\ref{2.91x}) tells us that $y_j = 1 - \lambda_j^2$.
Indeed, making this change of variable in (\ref{AC5}), and making the substitutions
(\ref{3.25a}), reclaims
(\ref{2.1b}) in the case $\beta = 2$.

For $\beta = 2$ and general $n_1 \ge n_2$, $N \ge n_1 + n_2$, a result equivalent to
(\ref{AC5}) is derived in a recent work of Simon and Moustakas \cite{SM05}.
Motivated by a quantum dot problem with three leads, they decomposed the 
$N \times N$ scattering matrix $S$ into a $3 \times 3$ block structure
\begin{equation}\label{srt}
S = \left [ \begin{array}{ccc} r_{11} & t_{12} & t_{13} \\
t_{21} & r_{22} & t_{23} \\
t_{31} & t_{32} & t_{33} \end{array} \right ]
\end{equation}
where each $r_{ii}$ is $N_i \times N_i$ and $t_{ij}$ is $N_i \times  N_j$, with
$N = N_1 + N_2 + N_3$. For $S$ a random unitary matrix with Haar measure, it is shown
that the eigenvalues of $t_{12}^\dagger t_{12}$ have p.d.f.~given by (\ref{AC5}) with
$\beta = 2$, $n_1 = N$, $n_2 = N_2$. This is consistent with our result because the
distribution of $S$ is unchanged by interchanging rows and columns. The first and
second block columns in (\ref{srt}) can be interchanged, effectively giving the
decomposition (\ref{2.4bb}) with $t_{12} = A_{N_1 \times N_2}$.

We remark that for $y_j \ll 1$ the term involving $(1-y_j)$ in
(\ref{AC5}) can be ignored. The resulting p.d.f.~is then the
$x_j \ll 1$ limit of the singular
values p.d.f.~of the matrix product $X^\dagger X$, where $X$ is an $n_1 \times n_2$
random Gaussian matrix with real $(\beta = 1)$, complex $(\beta = 2$) or real
quanternion elements $(\beta = 4)$
(see e.g.~\cite{Fo02}). This is consistent with
a recent result of Jiang \cite{Ji05}, 
who quantifies the degree to which the entries of orthogonal
or unitary random matrices can be approximated by entries of Gaussian random matrices
with real or complex entries respectively. 

\section{Random projections}
\setcounter{equation}{0}
In a recent work Collins \cite{Co05} computed the distribution of certain
products of random orthogonal projections. Here this approach, suitably modified, 
will be used to derive (\ref{AC5}). Knowledge is required of the following
result \cite{Mu82,Fo02}

\begin{prop}\label{pJ}
Let $c$ and $d$ be $n_1 \times m$ and $n_2 \times m$ Gaussian random matrices,
all elements identically and independently distributed (i.i.d.)
where $n_2,n_2 \ge m$,
indexed by the parameter $\beta = 1,2$ or 4. The parameter $\beta$ specifies that
the elements are real $(\beta = 1)$, complex $(\beta = 2)$ or real quaternion
$(\beta = 4)$. Let $C = c^\dagger c$, $D = d^\dagger d$. The distribution of
$J := (C+D)^{-1/2} C (C + D)^{-1/2}$ is proportional to
\begin{equation}\label{JE}
(\det J)^{a \beta/2} (\det (I - J))^{b \beta / 2}
\end{equation}
with
$$
a = n_1 - m + 1 - 2/\beta, \qquad b = n_2 - m + 1 - 2/\beta.
$$
\end{prop}

The p.d.f.~(\ref{JE}) is referred to as the Jacobi ensemble (see e.g.~\cite{Fo02}).
To make use of Proposition \ref{pJ}, let $X$ be a $N \times m$, $N \ge m$ 
i.i.d.~Gaussian random matrix indexed by $\beta$. Define $Q_1$ to be the top $n \times N$
submatrix of an $N \times N$ unitary matrix indexed by $\beta$. Because
$Q_1 Q_1^\dagger = I_{n}$, we have that $Q_1 X$ is distributed as an
$n \times m$ i.i.d.~Gaussian random matrix indexed by $\beta$ (see 
\cite[Thm.~2.3.1]{GN99}). Let $Q_2$ be the bottom $(N-n) \times N$ submatrix of
$U$ so that
\begin{equation}
Q_1^\dagger Q_1 + Q_2^\dagger Q_2 = I_N.
\end{equation}
Because $Q_2 Q_2^\dagger = I_{N-n}$, $Q_2 X$ is distributed as an
$(N-n) \times m$ i.i.d.~Gaussian matrix indexed by $\beta$.

Now set
\begin{equation}\label{YZ}
Y = X^\dagger Q_1^\dagger Q_1 X, \qquad
Z = X^\dagger Q_2^\dagger Q_2 X.
\end{equation}
Denoting by $W_p^{(\beta)}(n)$ the (Wishart) distribution of the matrix product
$A^\dagger A$, where $A$ is an $n \times p$, $(n \ge p)$ i.i.d.~Gaussian random matrix
indexed by $\beta$, we see that the distribution of $Y$ is $W_m^{(\beta)}(n)$
while the distribution of $Z$ is $W_m^{(\beta)}(N-n)$. The result of 
Proposition \ref{pJ} tells us that the eigenvalue p.d.f.~of
\begin{equation}\label{XyX}
(Y + Z)^{-1/2} Y ( Y + Z)^{-1/2} = (X^\dagger X)^{-1/2} Y  (X^\dagger X)^{-1/2} 
\end{equation}
is given by the Jacobi ensemble (\ref{JE}) with $n_1 = n$, $n_2 = N-n$.

According to the singular value decomposition, we can write
\begin{equation}\label{XyX1}
X = U_1 \Lambda U_2
\end{equation}
for $U_1$ an $N \times N$ unitary matrix, $U_2$ an $m \times m$ unitary matrix, and
$\Lambda$ an $N \times m$ diagonal matrix, with diagonal entries equal to the
positive square root of the eigenvalues of $X^\dagger X$. It follows from
(\ref{XyX1}) that
\begin{equation}\label{XyX2}
(X^\dagger X)^{1/2} = U_2 [\Lambda]_{m,  m} U_2
\end{equation}
where $[\Lambda]_{m,m}$ refers to the top $m \times m$ sub-block of
$\Lambda$, while (\ref{XyX2}) and (\ref{XyX1}) together imply
\begin{equation}\label{XyX3}
X (X^\dagger X)^{-1/2} = [U_1]_{N,m}   U_2^\dagger.
\end{equation}

Substituting for $Y$ in (\ref{XyX}) according to (\ref{YZ}), then using
(\ref{XyX3}) to substitute for $X (X^\dagger X)^{-1/2}$ tells us that the
distribution of
\begin{equation}\label{XyX4}
U_2 [U_1]_{N,m}^\dagger Q_1^\dagger Q_1 [U_1]_{N,m} U_2^\dagger
\end{equation}
is given by (\ref{JE}). We note
$$
U_1^\dagger  [U_1]_{N,m} = {\rm diag} (1,\dots,1,0,\dots,0)
$$
where there are $m$ $1$'s. Since the distribution of $Q_1$ is unchanged upon
multiplication by a unitary matrix, the distribution of (\ref{XyX4}) is the
same as the distribution of 
\begin{equation}
U_2 [Q_1^\dagger Q_1 ]_{m,m}  U_2^\dagger.
\end{equation}
In terms of the decomposition (\ref{2.4bb}) this in turn has the same
distribution as
\begin{equation}\label{XyX5}
A^\dagger A \Big |_{n_1 \mapsto n \atop n_2 \mapsto m}.
\end{equation}
Hence we have the distribution of (\ref{XyX5}) is given by
(\ref{JE}) with
$$
a = n - m +1 - 2/\beta, \qquad b = N - n - m +1 - 2/\beta,
$$
which is in precise agreement with (\ref{AC4}) upon the replacements
$n_1 \mapsto n$, $n_2 \mapsto m$ in the latter.

\section*{Acknowledgement}
This work was supported by the Australian Research Council.


\end{document}